\documentclass[twocolumn,showpacs,preprintnumbers,amsmath,amssymb,aps,prb]{revtex4}
\usepackage{graphicx}
\begin{document}

\title{Reversible Vector Ratchets for Skyrmion Systems}   
\author{
X. Ma$^{1,2}$, 
C. J. Olson Reichhardt$^{1}$, and  C. Reichhardt$^{1}$ 
} 
\affiliation{
$^1$ Theoretical Division,
Los Alamos National Laboratory, Los Alamos, New Mexico 87545 USA\\ 
$^2$ Department of Physics, University of Notre Dame, Notre Dame, Indiana 46556 USA\\
} 

\date{\today}
\begin{abstract}

  We show that ac driven skyrmions interacting with an asymmetric substrate
 provide a realization of a new class of ratchet system which we call a vector ratchet 
that arises due to the effect of the Magnus term on the skyrmion dynamics.
In a vector ratchet, the dc motion induced by the ac drive
can be described as a vector that can be rotated clockwise or counterclockwise
relative to the substrate asymmetry direction.
Up to a full $360^{\circ}$ rotation is possible  for varied ac amplitudes
or skyrmion densities.
In contrast to overdamped systems, in which ratchet motion is always parallel
to the substrate asymmetry direction, 
vector ratchets allow the ratchet motion to be in 
any direction relative to the substrate asymmetry.
It is also possible to obtain a reversal
in the direction of rotation of the vector ratchet, permitting the creation of a
reversible vector ratchet.
We examine
vector ratchets for ac drives applied parallel or perpendicular to the substrate
asymmetry direction,
and show that reverse ratchet motion 
can be produced by collective effects.
No reversals occur for an isolated skyrmion on
an asymmetric substrate.
Since a vector ratchet
can produce motion in any direction,
it could represent a new method for controlling skyrmion motion for spintronic applications.
\end{abstract}
\maketitle

\section{Introduction}

In a rocking ratchet, a particle or collection of particles
interacting with an asymmetric substrate undergoes a net dc drift when
subjected to an ac drive \cite{1,2}, as observed for
vortices in type-II superconductors interacting with
one-dimensional (1D) \cite{3,4,5}
or two-dimensional (2D) asymmetric substrates \cite{4w,5w,6,7}.
In the single particle limit,
the ratchet motion is typically in the
easy flow direction of the substrate asymmetry;
however, when collective effects come into play,
it is possible for a reverse ratchet effect to occur in which
the particles move along the opposite or hard flow direction of the substrate
asymmetry.  Reversals of the ratchet direction
can occur when parameters such as
the ac amplitude, particle density,
or substrate strength are varied \cite{1,2,8,9,10,11,12,13}.
It is also possible to observe a transverse ratchet effect in which
the net dc drift of the particles is perpendicular to applied ac drive.
For such transverse ratchets,
when the ac drive is applied transverse to the substrate asymmetry direction,
the resulting dc drift
is parallel to the substrate asymmetry in either the easy or hard flow direction
\cite{9,15,16,17}.

In many of the experimentally studied systems where
ratchet effects occur, such as vortices in type-II superconductors \cite{8,10,12,13,14}
or colloids \cite{20,21}, the motion of the particles is effectively overdamped.
Recently a new particlelike excitation called skyrmions was discovered
in chiral magnets \cite{22,23,24}.  These
skyrmions have many similarities to vortices in type-II superconductors in that they
exhibit particlelike properties and
have a mutually repulsive interaction that leads to the
formation of a triangular skyrmion lattice \cite{22,23}.
Skyrmions can be driven with an applied current \cite{24,New,25,26,27,28,29}
and exhibit pinning-depinning phenomena \cite{25,27,29}.
A key difference between superconducting vortex and skyrmion systems is that in
addition to the damping,
skyrmion motion involves a strong non-dissipative Magnus effect
which rotates the skyrmion velocity
into the direction perpendicular to the net applied external forces.
This Magnus term can be ten or more times larger than
the damping term \cite{24,25,27,30}.
In the absence of pinning,
under a dc drive the Magnus effect causes the
skyrmions to move at an angle, the skyrmion Hall angle $\theta_{Sk}$,
with respect to the driving direction, where
$\theta_{Sk} \sim \tan^{-1}(\alpha_{m}/\alpha_{d})$ and $\alpha_{m}/\alpha_{d}$ is the ratio
of the Magnus term to the damping term.
In the presence of pinning, the skyrmion Hall angle
has a strong drive dependence \cite{31,32,33,34,litzius}.
Skyrmions have now been stabilized at room
temperature \cite{29,35,36}
making them promising candidates for a variety of spintronic
applications \cite{37}, any of which would require
the ability to precisely control the skyrmion motion.
One method for achieving such control would be to exploit ratchet effects.

In previous numerical work, it was shown that an individual skyrmion in
a 2D system interacting with a quasi-1D asymmetric substrate exhibits
a rocking ratchet effect when the ac drive is applied along the substrate
asymmetry direction \cite{38}.
In this case, the resulting dc skyrmion velocity has components both parallel and
perpendicular to the substrate asymmetry direction
due to the Magnus term.
A new type of ratchet effect, called a Magnus ratchet, was shown to occur
when the ac drive is applied perpendicular to the substrate asymmetry direction
\cite{38}.
Here, 
the Magnus term induces skyrmion velocity components both parallel and perpendicular
to the ac drive.  As a result, the skyrmions translate partially
along the substrate asymmetry direction, permitting ratcheting motion
to occur.
In the overdamped limit, this Magnus ratchet effect is lost.
In the single skyrmion limit for both longitudinal and transverse ac driving,
the ratchet flux is always aligned with
the easy flow direction of the substrate asymmetry,
so an open question is whether it is possible to
realize a reversible skyrmion ratchet effect.

In this work we consider skyrmions driven by ac forces over
gradient pinning arrays.
Previous studies of such arrays in the overdamped limit for superconducting vortices
demonstrated that both longitudinal and transverse ratchet effects
as well as ratchet reversals
occur as a function of ac amplitude and vortex density \cite{17,39}.
Here we  show that for ac drives applied either parallel or perpendicular to the
substrate asymmetry direction,
when a finite Magnus term is present, ratchet effects occur even in regimes where there
is no ratchet motion in the overdamped limit,
while multiple reversals of the ratchet effect can appear when
the ac amplitude, the skyrmion density, or
the ratio $\alpha_m/\alpha_d$ of the Magnus term to the damping term
is varied.
The net dc drift of the skyrmions can be described as a vector which
contains information about the magnitude of the drift and the angle between the drift
direction and
the substrate easy flow direction.
With changing
$\alpha_{m}/\alpha_{d}$, ac amplitude, or skyrmion density,
the ratchet vector undergoes either a clockwise or
counterclockwise rotation of up to 360$^\circ$,
indicating that
ratcheting motion can occur in any direction
for a 2D system.
It is even possible to have a reversal in the direction of rotation of the ratchet vector.
This system thus represents a new class of ratchet which we call a vector ratchet,
and we predict that
vector ratchets should be a general feature of any
system in which
Magnus effects are important,
including skyrmions in chiral magnets \cite{24}, skyrmion phases
in p-wave superconductors \cite{40,41,42}, rotating colloids \cite{43},  and
charged particles in magnetic fields such as dusty plasmas \cite{44,45}.
Additionally, since vector ratchets allow for motion
in any direction, they also could serve as a new method
to control skyrmion motion for spintronic applications.

\begin{figure}
\includegraphics[width=3.5in]{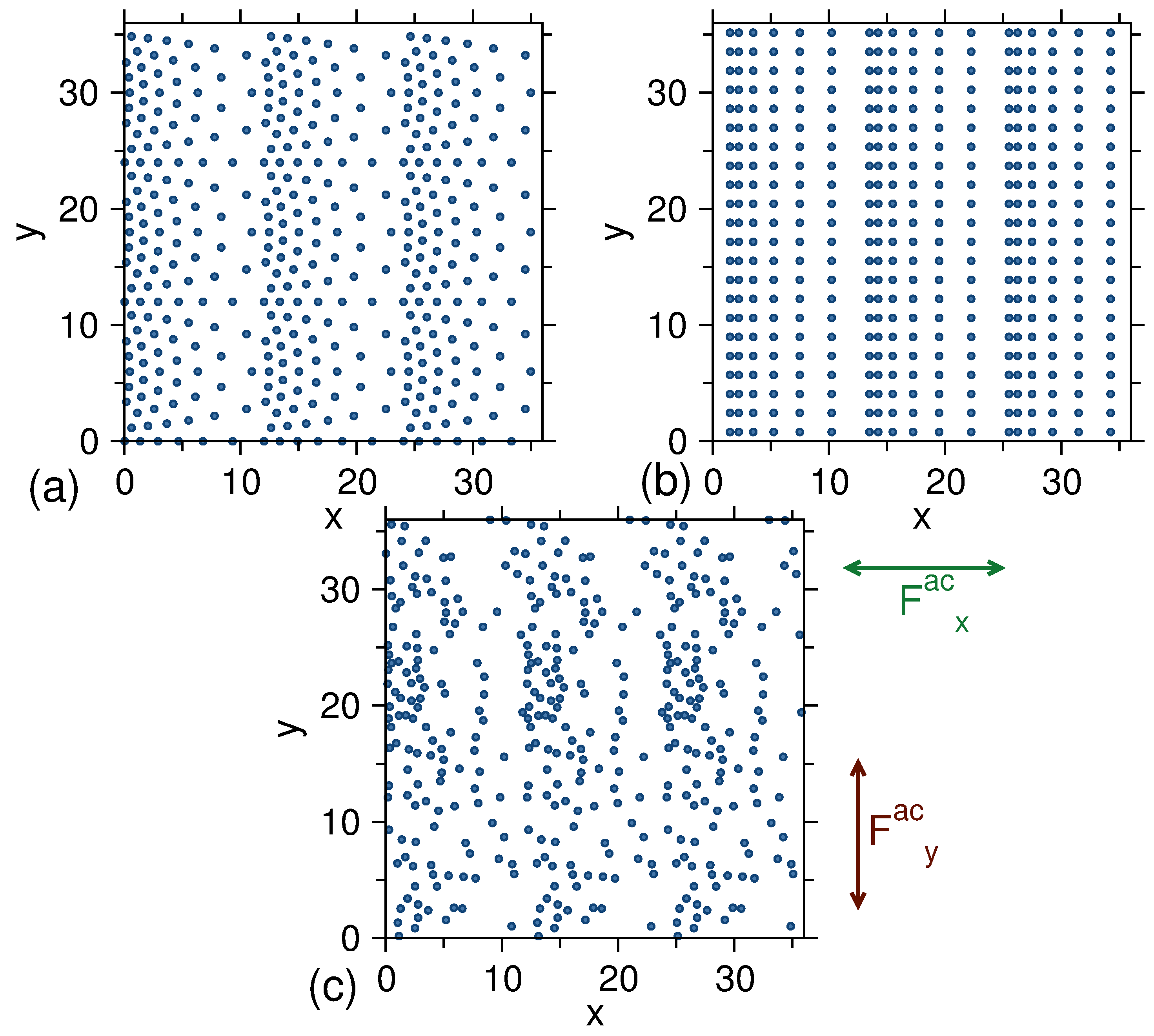}
\caption{ Circles: Pinning site locations.
  (a) Conformal gradient array. (b) Square gradient array. (c)
  Random gradient array.
  Green arrow: direction of longitudinal drive $F^{ac}_x$.  Red arrow:
  direction of transverse drive $F^{ac}_y$.
}
\label{fig:1}
\end{figure}

\section{Simulation}   

We model a 2D system of size $L \times L$ with periodic boundary conditions in
the $x$- and $y$-directions containing $N_{s}$ skyrmions
at a density of $\rho_{s} = N_{s}/L^2$.
We place $N_{p}$ pinning sites in one of the periodic gradient array configurations
illustrated in Fig.~\ref{fig:1}.
We focus
primarily on the conformal array shown in Fig.~\ref{fig:1}(a),
which is produced by performing
a conformal transformation on a uniform triangular array of
pinning sites, as described in
detail in previous work on pinning \cite{46,47} and ratchet effects \cite{17,39} for
superconducting vortices in conformal pinning arrays.
Successful experimental realizations of 
conformal pinning arrays
for superconducting vortex systems \cite{48,49}
suggest that  similar nanofabrication techniques could be used to
create such arrays for skyrmion systems.
Figure~\ref{fig:1}(b) illustrates the square gradient array, produced by
subjecting a square pinning lattice to a gradient along the $x$ direction, while
Fig.~\ref{fig:1}(c) shows the random gradient array, generated by introducing the same
$x$ direction pinning density gradient to a random pinning array.
We apply an ac driving force to the skyrmions
of either $F^{ac}_x$, in the longitudinal or $x$ direction, or $F^{ac}_y$, in the
transverse or $y$ direction,
and measure the average net displacement of the skyrmions
as a function of ac cycle.

To simulate the skyrmion motion we use a
modified Theile equation \cite{50} described in Refs.~\cite{30,32,33}
that takes into account skyrmion-skyrmion interactions
and skyrmion-pinning interactions.
The
equation of motion of a single skyrmion $i$ is
\begin{equation}
\alpha_{d}{\bf v}_{i} + \alpha_{m}{\hat z}\times {\bf v}_{i} = {\bf F}^{ss}_{i} + {\bf F}^{sp}_{i} + {\bf F}^{ac} .
\end{equation}
Here
${\bf r}_{i}$ is the location of skyrmion $i$ and 
${\bf v}_{i} = d{\bf r}_{i}/dt$ is the 
skyrmion velocity.
The damping term with prefactor $\alpha_{d}$
generates a skyrmion velocity component in the direction of
the net external forces, while the Magnus term with prefactor
$\alpha_{m}$ generates 
a skyrmion velocity component perpendicular to the net external force direction.
The repulsive skyrmion-skyrmion interactions are given by
${\bf F}^{ss}_{i} = \sum^{N_{s}}_{j=1}K_{1}(R_{ij}){{\bf \hat r}}$,
where $R_{ij} = |{\bf r}_{i} - {\bf r}_{j}|$ is the distance between
skyrmions $i$ and $j$, and
$K_{1}$ is the modified Bessel function
which falls off exponentially for large $R_{ij}$.
The pinning force ${\bf F}^{sp}_{i}$ is modeled as arising from
attractive nonoverlapping harmonic traps
of radius $R_{p}$ which can exert a maximum pinning
force of $F_{p}$.
The ac driving force
is ${\bf F}^{ac} = F^{ac}_{\beta}\sin(\omega t){\hat {\bf \beta}}$,
where $\beta=x$ for longitudinal driving and
$\beta=y$ for transverse driving, as shown schematically in Fig.~\ref{fig:1}.
To characterize the ratchet effect, we measure
the average net displacement of the skyrmions over time in
both the $x$ and $y$ directions to obtain
$\langle \Delta X\rangle = N_{s}^{-1}\sum^{N_{s}}_{i=1}(x_i(t) - x_{i}(t_{0}))$ and
$\langle\Delta Y\rangle = N_{s}^{-1}\sum^{N_{s}}_{i=1}(y_i(t) - y_i(t_{0}))$,
where $(x_i,y_i)(t)$ is the position of skyrmion $i$ at time
$t$ and $t_{0}$ is the initial reference time.
We use a measurement interval of $t-t_0=400$ ac drive cycles, and the initial
reference time $t_0$ is taken to be no less than 50 ac drive cycles after the system is
initialized.
The system size $L=36$ and the spacing between repeated tilings of our gradient
pinning arrays is
$a_{p} = 12$.  The average spacing between individual pinning sites is
$a = 1.82$.
In this work we focus on samples with skyrmion density $n_{s}=0.3$,
filling fraction of $n_{s}/n_{p} = 1.0$,
pinning radius of $R_{p} = 0.3$, and
pinning force of $F_{p} = 0.1$.

\section{dc Depinning}

\begin{figure}
\includegraphics[width=3.5in]{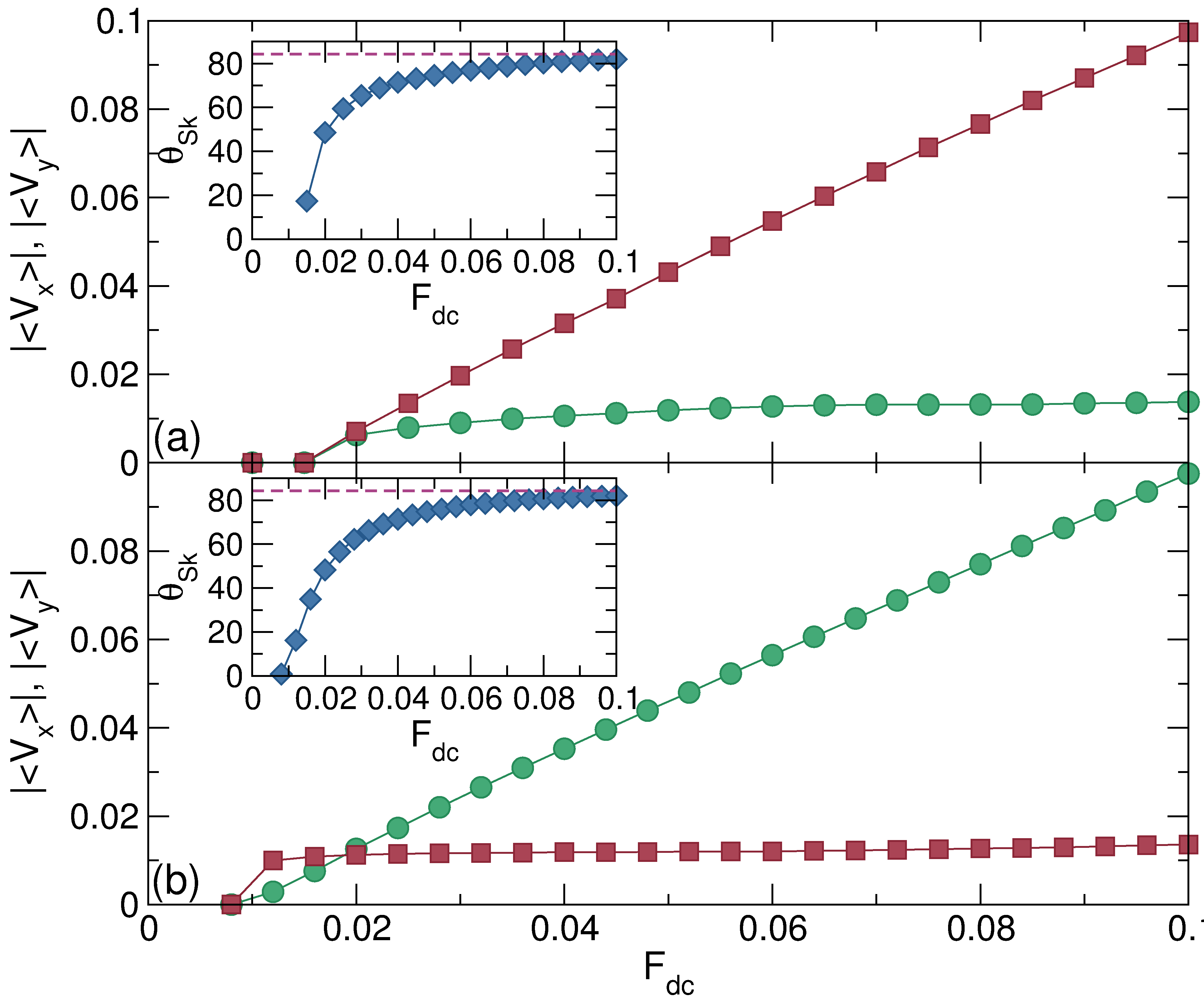}
\caption{The velocity-force curves
  $|\langle V_x\rangle|$ (green circles) and
  $|\langle V_y\rangle|$ (red squares) vs dc drive $F_{dc}$ for the conformal pinning array
  in Fig.~\ref{fig:1}(a) with
  $\alpha_{m}/\alpha_{d} = 9.962$.
  (a) For dc driving in the positive $x$ direction,
  the critical depinning threshold is $F_{c} \approx 0.015$.
  Inset: The skyrmion Hall angle
    $\theta_{Sk} = \tan^{-1}(|\langle V_{\perp}\rangle|/|\langle V_{||}\rangle|)$ vs dc drive
  amplitude $F_{dc}$, where $V_{\perp}=V_y$ and $V_{||}=V_x$.
  The dashed line indicates the
  pin free limit of $\theta_{Sk}=84.267^{\circ}$.
  (b) For dc driving in the positive $y$ direction, close to
  the depinning threshold the skyrmion motion is strongly guided along the $y$ direction
  by the pinning sites.
  Inset: $\theta_{Sk}$ vs $F_{dc}$,
  where $V_{\perp}=V_x$ and $V_{||}=V_y$, shows that $\theta_{Sk}$ is nearly zero
  at low drives and increases
  with increasing $F_{dc}$.
  The dashed line indicates the
  pin free limit of $\theta_{Sk}=84.267^{\circ}$.
}
\label{fig:2}
\end{figure}

We first apply a dc drive to the conformal pinning array
sample in order to determine the depinning threshold.
In Fig.~\ref{fig:2}(a)  
we plot $|\langle V_{x}\rangle|$ and $|\langle V_{y}\rangle|$ versus
the dc drive amplitude $F_{dc}$ for driving in the positive $x$-direction
in a sample with $\alpha_{m}/\alpha_{d} = 9.962$.
The inset shows the
skyrmion Hall angle
$\theta_{Sk} = \tan^{-1}(|\langle V_{y}\rangle|/|\langle V_{x}\rangle|)$
$\theta_{Sk} = \tan^{-1}(|\langle V_{\perp}\rangle|/|\langle V_{||}\rangle|)$
versus $F_{dc}$,
where $V_{\perp}=V_y$ and $V_{||}=V_x$.
The depinning threshold $F_c$ is close to $F_{c} = 0.015$.
The Hall angle
$\theta_{Sk} \approx 20^{\circ}$ at low drives, and
gradually increases with increasing $F_{dc}$ until it reaches the
expected pin-free value of $\theta_{Sk} = 84.267^{\circ}$.
This
strong dependence of the skyrmion Hall angle        
on the external drive in the presence of pinning
was observed in previous studies of particle-based
\cite{32,33} and continuum-based \cite{31} simulations as well as in
experiments \cite{34}.
For dc driving in the positive $y$ direction,
Fig.~\ref{fig:2}(b) shows
that the depinning threshold has a lower value of $F_{c} = 0.01$.
Near depinning, there is a stronger guiding effect
in the $y$-direction as the skyrmions move through the low pinning density region
of the conformal array.  As a result, the motion just above depinning
is almost completely locked in the $y$ direction, giving a Hall angle close to
zero, as shown in the inset of Fig.~\ref{fig:2}(b).

\section{Ratchet Effects with Longitudinal and Transverse AC Drives}

\begin{figure}
\includegraphics[width=3.5in]{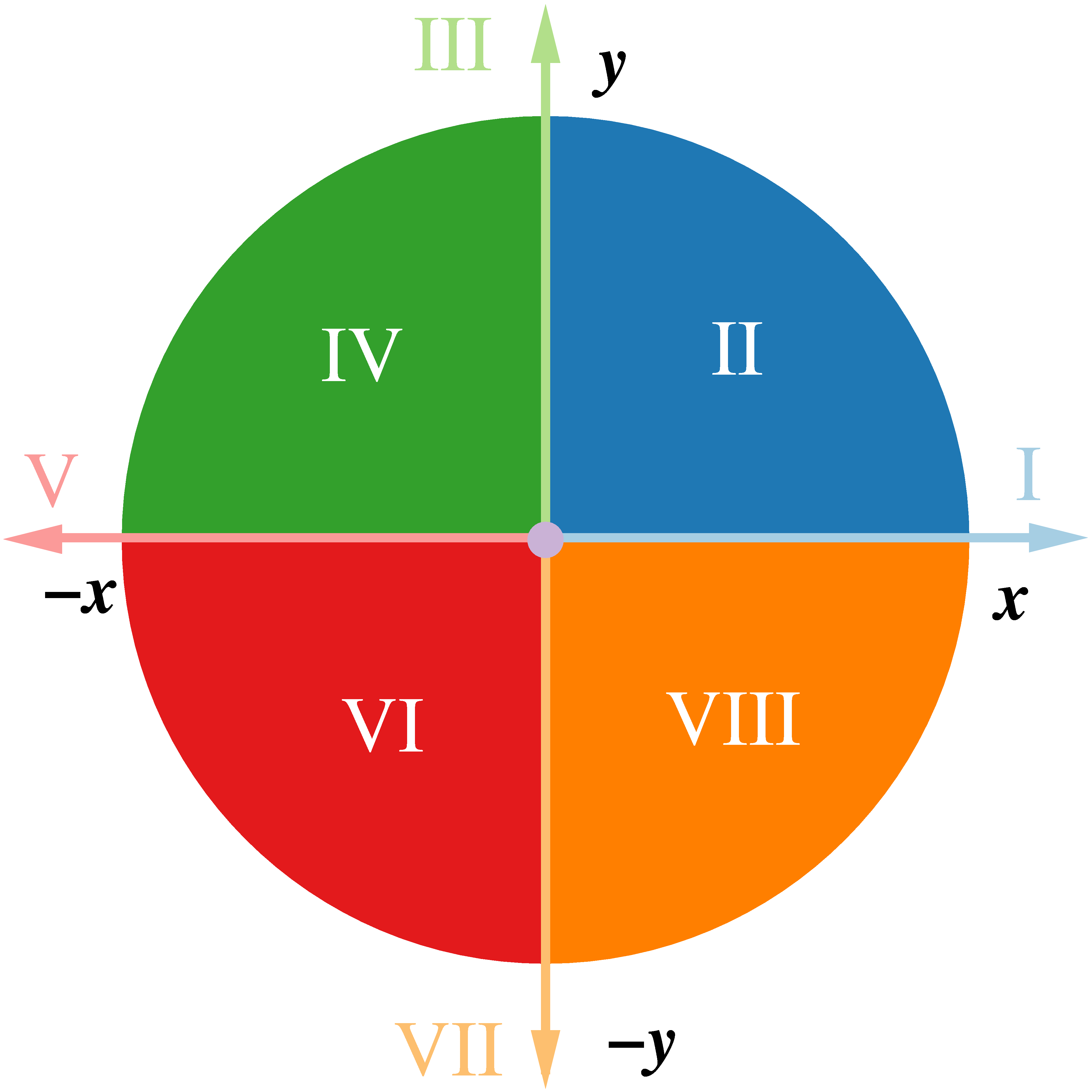}
\caption{A diagram showing the
  eight possible types of vector ratchet motion
  for the conformal pinning array in Fig.~\ref{fig:1}(a).
  The net dc drift in the $(x,y)$ direction for each type is:
  I: $(+x, 0)$ (light blue);
  II: $(+x,+y)$ (dark blue);
  III: $(0, +y)$ (light green);
  IV: $( -x, +y)$ (dark green);
  V: $(-x, 0)$ (pink);
  VI: $(-x, -y)$ (red);
  VII $(0, -y)$ (light orange); and
  VIII $(+x, -y)$ (dark orange).
  In addition, we define type IX with $(0,0)$ (purple) to be a state with
  no ratcheting motion.
}
\label{fig:3}
\end{figure}

To analyze the ratchet effect, we apply an ac drive to the conformal pinning array sample
in Fig.~\ref{fig:1}(a)
along the longitudinal ($F^{ac}_x$) or transverse ($F^{ac}_y$) direction,
as indicated
by the arrows in Fig.~\ref{fig:1}.
In the overdamped case,
only two types of ratchet effects occur:
a net dc motion along the positive or negative $x$ direction, parallel to the drive,
for longitudinal driving, and
a net dc motion along the positive or negative $x$ direction, perpendicular to the drive,
for transverse driving.
In contrast, there can be up to eight types of motion for
a Magnus induced ratchet.
As shown in Fig.~\ref{fig:3}, these are
type I, with net motion in the positive $x$ direction only;
type II, with net motion in the positive $x$ and positive $y$ directions;
type III, with net motion in the positive $y$ direction only;
type IV, with net motion in the negative $x$ and positive $y$ directions;
type V, with net motion in the negative $x$ direction only;
type VI, with net motion in the negative $x$ and negative $y$ directions;
type VII, with net motion in the negative $y$ direction only;
and type VIII, with net motion in the positive $x$ and negative $y$ directions.
We also refer to type IX, where there is no net motion in either direction, indicating the
lack of a ratchet effect.
Overdamped systems exhibit ratchet types I and V.

\begin{figure}
\includegraphics[width=3.5in]{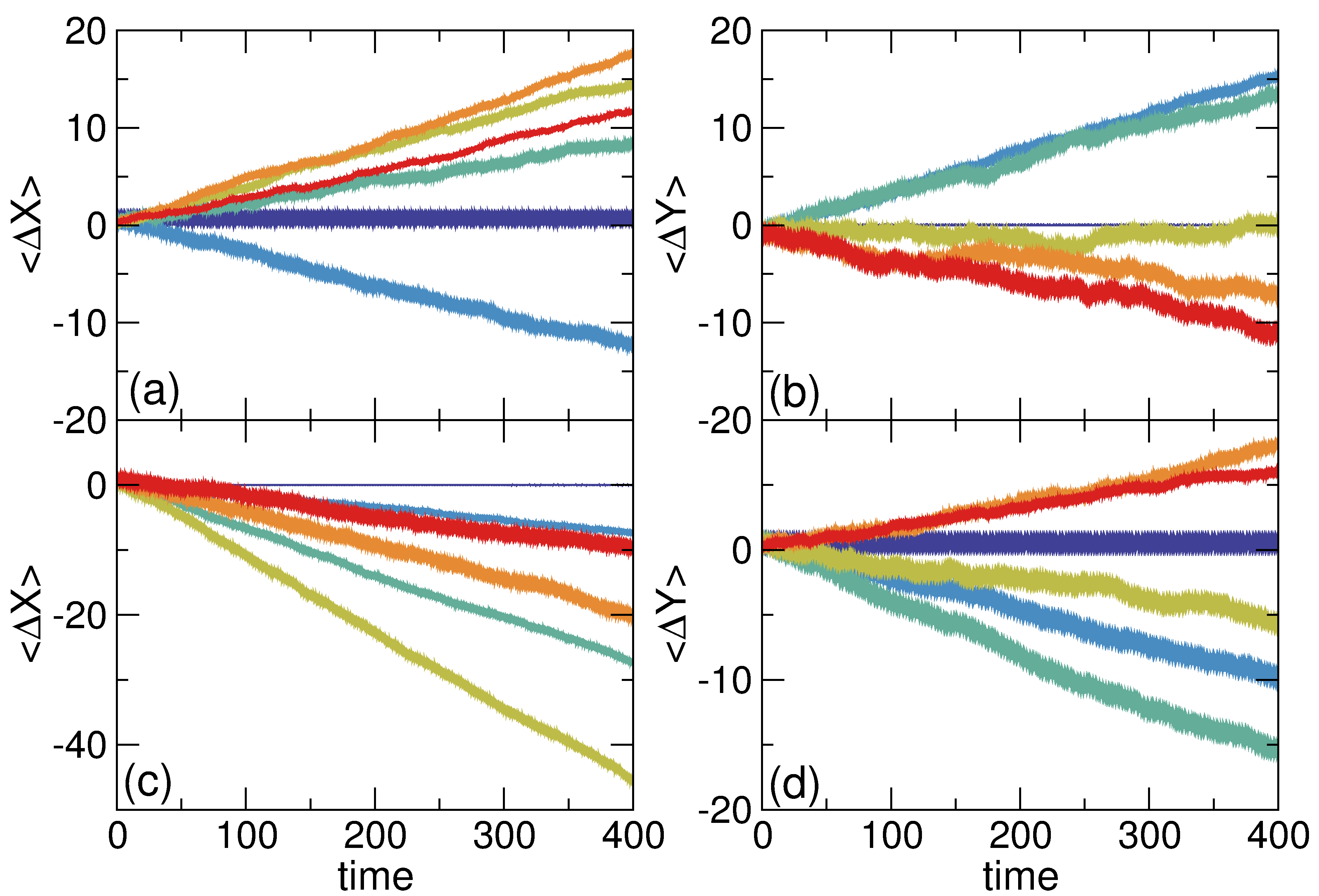}
\caption{(a,b) The average cumulative skyrmion displacement
  $\langle \Delta X\rangle$ (a) and $\langle \Delta Y\rangle$ (b) vs
  time in ac cycles
  for the conformal pinning array under longitudinal ac driving
  with $F^{ac}_x = 0.04$ at
  $\alpha_{m}/\alpha_{d} = 0$ (dark blue), 1.36 (light blue), 4.0 (dark green),
  8.0 (light green), 10 (orange), and 20 (red).
  There is no ratchet motion when $\alpha_{m}/\alpha_{d} = 0$, but for
  $\alpha_m/\alpha_d \neq 0$, we observe ratchet reversals in both the $x$ and $y$
  directions.
  (c,d) $\langle \Delta X\rangle$ (c) and $\langle \Delta Y\rangle$ (d) vs time in
  ac cycles
  for the same system for  transverse ac driving
  at $F^{ac}_y=0.04$ and $\alpha_m/\alpha_d=0$ (dark blue),
  1.2 (light blue), 1.6 (dark green), 2.6 (light green), 10 (orange), and 20 (red).
  In this case the ratchet motion for $\alpha_m/\alpha_d \neq 0$
  is always in the negative $x$ direction and shows a reversal in the $y$ direction.
}
\label{fig:4}
\end{figure}

We now consider a case where there is no ratchet effect
in the overdamped limit for either longitudinal or transverse ac driving, and we
vary the ratio $\alpha_m/\alpha_d$ of the Magnus term to the damping term.
In Fig.~\ref{fig:4}(a,b) we plot the average
cumulative displacement per skyrmion
$\langle \Delta X\rangle$  in the $x$ direction
and $\langle \Delta Y\rangle$ in the $y$ direction
versus time in ac cycles
for a system with $F^{ac}_x = 0.04$ in the longitudinal or $x$ direction.
At $\alpha_{m}/\alpha_{d} = 0$,
$\langle \Delta X\rangle =  0$ and $\langle \Delta Y\rangle = 0$,
indicating the absence of a ratchet effect.
For
$\alpha_{m}/\alpha_{d} = 1.36$, the skyrmions move
in the negative $x$ direction and the positive $y$ direction, which in the
notation of Fig.~\ref{fig:3} is a type IV ratchet.
The negative $x$ direction is the easy flow direction of the
substrate asymmetry.
As $\alpha_m/\alpha_d$ increases from 4 to 20, a reversal of the ratchet effect
occurs in which $\langle \Delta X\rangle$ becomes positive
so that the skyrmions are moving in the hard flow direction of the
substrate asymmetry.
The corresponding $\langle \Delta Y\rangle$ remains in the positive
$y$ direction for $\alpha_{m}/\alpha_{d} = 4$,
resulting in a type II ratchet,
while for
$\alpha_{m}/\alpha_{d} = 8.0$, $\langle \Delta X\rangle > 0$
and $\langle \Delta Y\rangle = 0$,
giving a type I ratchet.
For  $\alpha_{m}/\alpha_{d} = 10$ and $20$,
there is a $y$ direction reversal with
$\langle \Delta Y\rangle < 0$ and $\langle \Delta X\rangle > 0$,
producing a type VIII ratchet.
The sequence of ratchet types that appear
as a function of increasing $\alpha_{m}/\alpha_{d}$, including the lack of a ratchet
effect at $\alpha_m/\alpha_d=0$, 
is IX-IV-III-II-I-VIII, so that the ratchet direction is moving clockwise around
the diagram in Fig.~\ref{fig:3}.

In Fig.~\ref{fig:4}(c,d) we show $\langle \Delta X\rangle$ and $\langle \Delta Y\rangle$
versus time in ac cycles for transverse or $y$ direction ac driving with $F^{ac}_y=0.04$,
where there is again no ratchet effect for
$\alpha_{m}/\alpha_{d} = 0.$
We find that $\langle \Delta X\rangle$ is always negative
but that there is a reversal in $\langle \Delta Y\rangle$, which is negative for
$0 < \alpha_{m}/\alpha_{d} < 10$, giving a type VI ratchet, and
positive for
$\alpha_{m}/\alpha_{d} \geq 10$,  producing a type IV ratchet.
The ratchet
sequence in this case is
IX-VI-V-IV.
The maximum ratchet flow magnitude
is $3.75$ times larger for transverse ac driving than for longitudinal ac driving.

\begin{figure}
\includegraphics[width=3.5in]{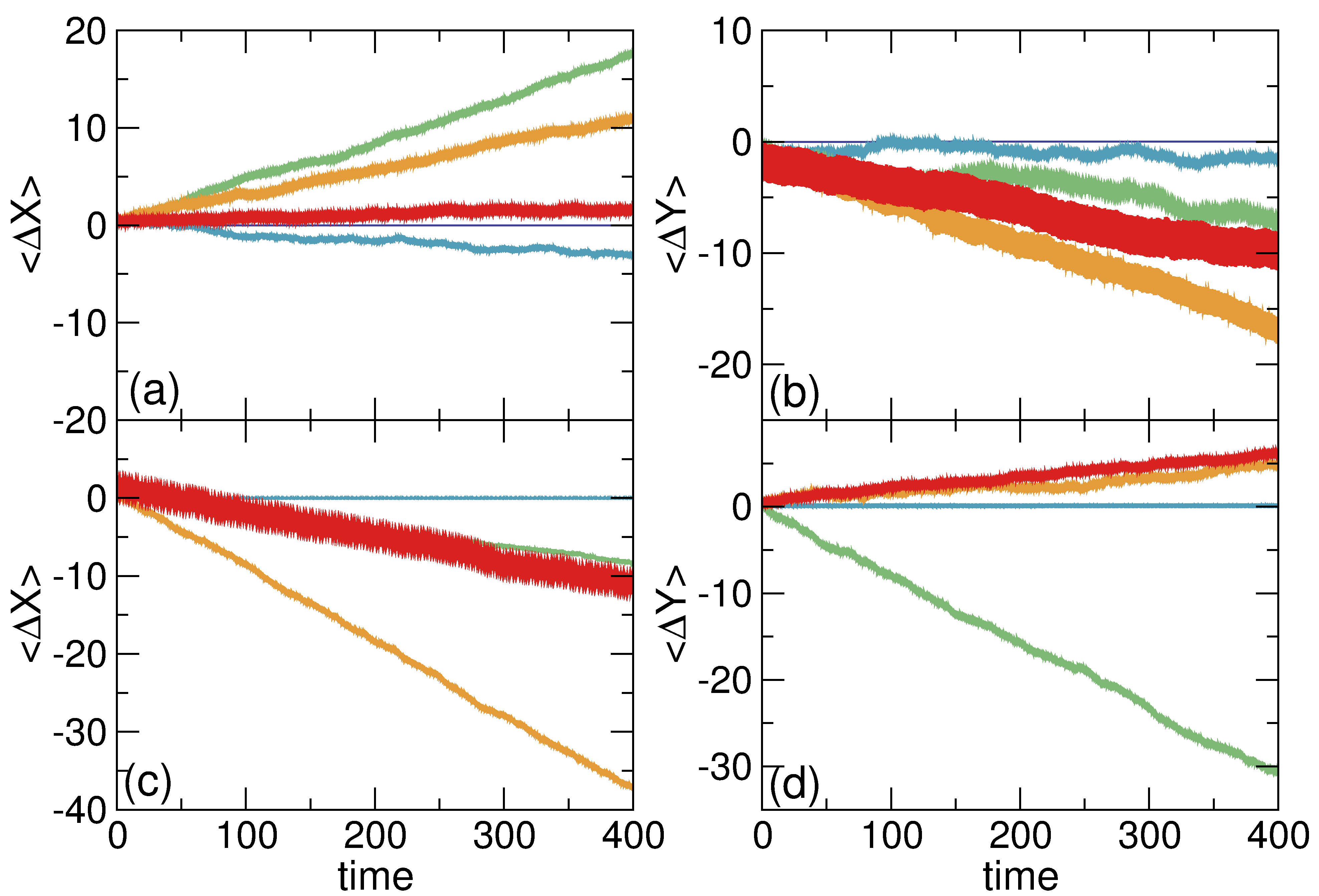}
\caption{ (a,b) $\langle \Delta X\rangle$ (a) 
  and $\langle \Delta Y\rangle$ (b) vs time in ac cycles
  for the conformal
  pinning array for
  longitudinal ac driving with
  $\alpha_{m}/\alpha_{d} = 10$
  at $F^{ac}_x = 0$ (dark blue), 0.025 (light blue), 0.04 (green), 0.06 (orange), and
  $0.08$ (red).
  There is no ratchet motion at $F^{ac}_x = 0$, but for $F^{ac}_x>0$, ratchet reversals
  occur in both the $x$ and $y$ directions.
  (c,d) $\langle \Delta X\rangle$ (c)
  and $\langle \Delta Y\rangle$ (d) for the same system for
  transverse ac driving
  at $\alpha_m/\alpha_d=10$ and
  $F^{ac}_y = 0$ (dark blue), 0.007 (light blue), 0.015 (green), 0.021 (orange),
  and $0.06$ (red).  The
  ratchet effect is always in the negative $x$ direction
  and shows a reversal in the $y$ direction.
}
\label{fig:5}
\end{figure}

We also observe ratchet reversals at
fixed $\alpha_{m}/\alpha_{d} = 9.962$ as we vary $F^{ac}_x$, as shown in
Fig.~\ref{fig:5}(a,b) where we plot $\langle \Delta X\rangle$ and
$\langle \Delta Y\rangle$ versus time in ac cycles.
At $F^{ac}_x = 0$ there is no ratchet effect, while
at $F^{ac}_x = 0.025$, there is a weak
ratchet effect in the negative $x$ direction that crosses over to
a positive $x$ ratchet for
$F^{ac}_x = 0.04$ and $0.06$.
The ratchet effect in the $y$-direction is always negative.
At $F^{ac}_x = 0.08$, the motion is predominately
in the negative $y$ direction  with almost no $x$ direction movement,
so the resulting sequence of ratchet types is IX-V-VIII-VI.

In Fig.~\ref{fig:5}(c,d) we show $\langle \Delta X\rangle$ and $\langle \Delta Y\rangle$
versus time
for the $\alpha_m/\alpha_d=9.962$ sample under transverse ac driving.
The ratchet effect is
always in the negative $x$ direction,
with the largest ratchet flow occurring at
$F^{ac}_y = 0.021$, a drive at which a single skyrmion translates
a distance larger than the entire system length $L$ during
half of an ac drive cycle.
The ratchet motion transitions from weak to strong negative $y$ direction flow with
increasing $F^{ac}_y$ before switching to positive $y$ direction flow
for $F^{ac}_y>0.02$,
giving a ratchet sequence of 
IX-VI-V-IV.

\begin{figure}
\includegraphics[width=3.5in]{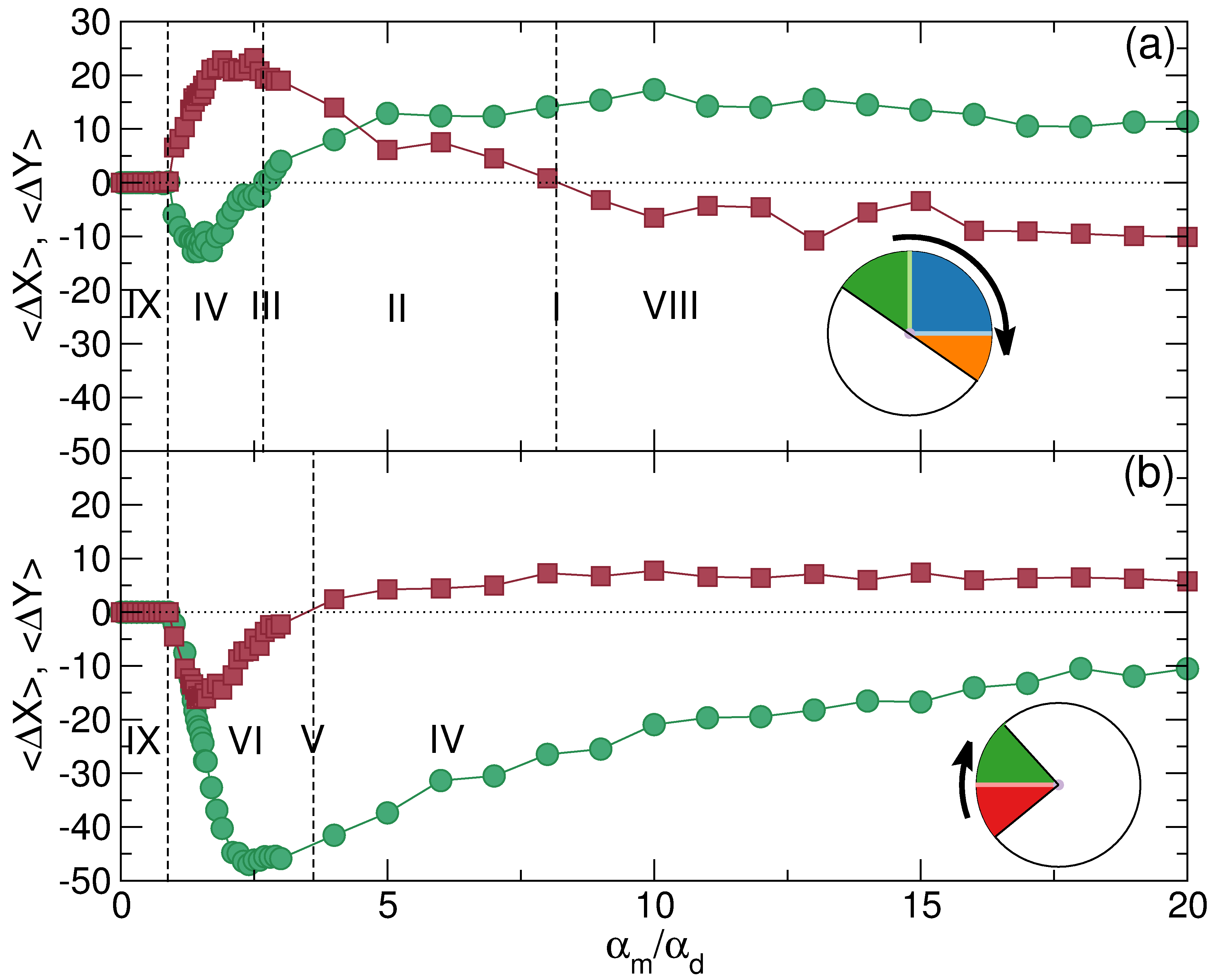}
\caption{The value of $\langle \Delta X\rangle$ (green circles)
  and $\langle \Delta Y\rangle$ (red squares) after
  400 ac cycles vs $\alpha_m/\alpha_d$ for the system in Fig.~\ref{fig:4}.
  (a) Driving in the $x$ direction with $F^{ac}_x=0.04$.
  The ratchet sequence is IX-IV-III-II-I-VIII,
  so that the flow rotates clockwise by $180^\circ$ as indicated in the inset, which is
  based on the schematic in Fig.~\ref{fig:3}.
  (b) Driving in the $y$ direction with $F^{ac}_y=0.04$.
  The ratchet sequence is IX-VI-V-IV giving a counterclockwise rotation of
  $90^\circ$ as shown in the inset.
}
\label{fig:6}
\end{figure}

In Fig.~\ref{fig:6}(a) we plot the values of
$\langle \Delta X\rangle$ and $\langle \Delta Y\rangle$ after 400 ac cycles
as a function of $\alpha_{m}/\alpha_{d}$
for the system in Fig.~\ref{fig:4}(a,b).
At $\alpha_{m}/\alpha_{d} = 0$ there is no ratchet effect, which we term a type IX
ratchet,
while for $0.75 < \alpha_{m}/\alpha_{d}  <  2.6$, the ratchet motion is in the negative
$x$ and positive $y$ directions, which is a type IV ratchet.
The ratchet motion passes through zero in the $x$ direction at $\alpha_{m}/\alpha_{d} = 2.6$
while continuing to flow in the positive $y$ direction,
giving a type III ratchet.  This is also an example of
a transverse ratchet effect in which a longitudinal dc drive produces drift
motion strictly in the transverse direction.
In the interval $2.6 < \alpha_m/\alpha_d<8.0$ we find a type II ratchet with
positive $x$ and positive $y$ motion, followed by a type I or strictly positive $x$ direction
ratchet at $\alpha_m/\alpha_d=8.0$.
Finally, for $\alpha_m/\alpha_d>8.0$, we observe type VIII flow with positive $x$ and
negative $y$ motion.
The sequence of ratchet types as a function of $\alpha_m/\alpha_d$ is indicated in
the inset of Fig.~\ref{fig:6}(a), where the flow begins in region IV and gradually
rotates clockwise by nearly $180^\circ$.
For driving in the $y$ direction,
Fig.~\ref{fig:6}(b) shows
that initially the system exhibits a type VI ratchet effect with negative $x$ and $y$
motion, passes through a type V ratchet in which motion occurs only in the negative $x$
direction despite the fact that the driving is applied along the $y$ direction,
and then finally enters a broad type IV ratchet region in which the flow is in
the negative $x$ and positive $y$ directions.
The flow sequence is thus IX-VI-V-IV, and the flow rotates clockwise in the
inset of Fig.~\ref{fig:6}(b) by about 90$^\circ$ as a function of $\alpha_m/\alpha_d$.

\begin{figure}
\includegraphics[width=3.5in]{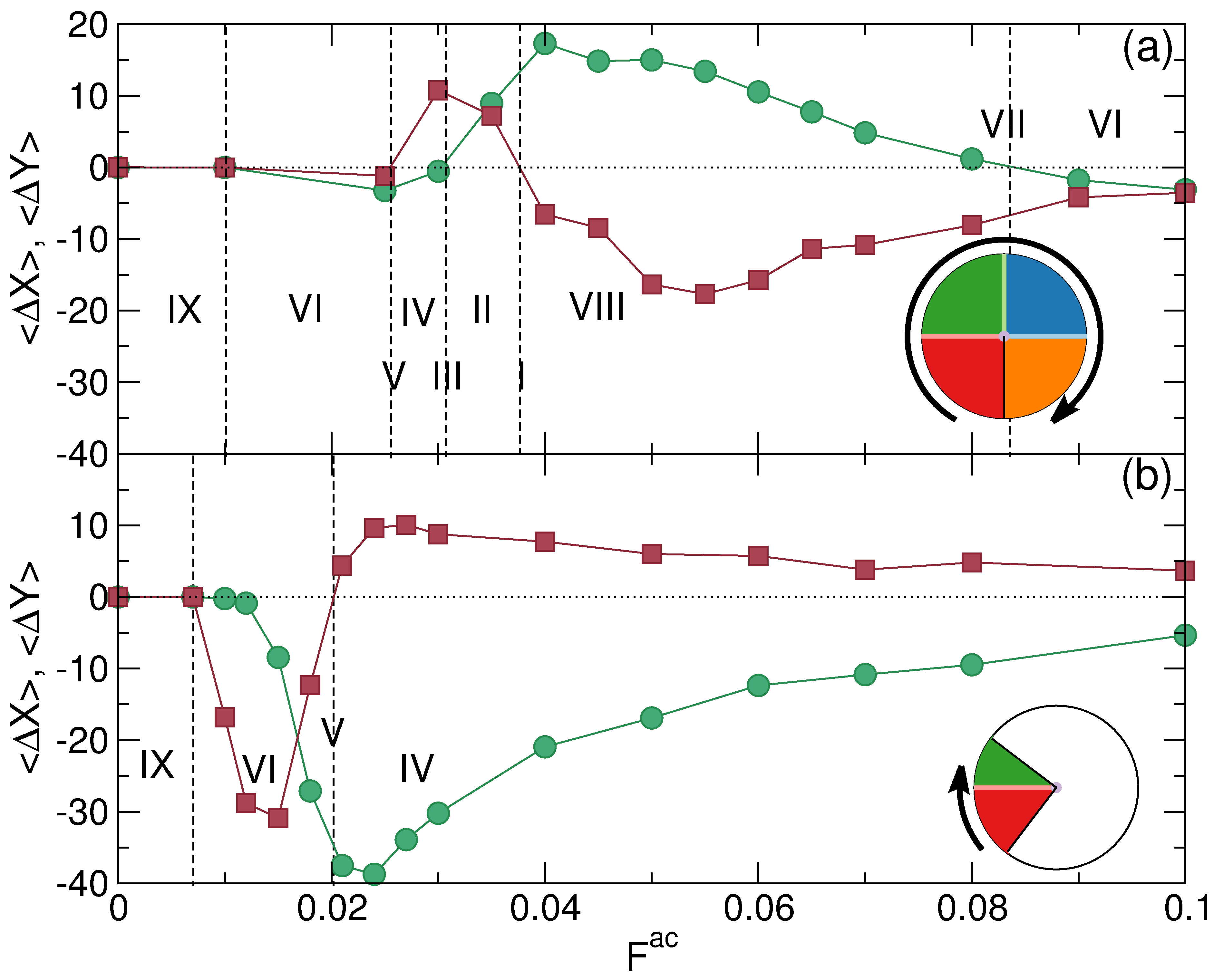}
\caption{
  $\langle \Delta X\rangle$ (green circles) and
  $\langle \Delta Y\rangle$ (red squares) after 400 ac cycles vs $F^{ac}$
  for the system in Fig.~\ref{fig:4}
  at $\alpha_{m}/\alpha_{d}=10.0$.
  (a) Driving in the $x$ direction with $F^{ac}_x$.
  The ratchet sequence is IX-VI-V-IV-III-II-I-VIII-VII-VI,
  giving a clockwise rotation of $360^\circ$ as indicated in the inset.
  (b) Driving in the $y$ direction with $F^{ac}_y$.
  The ratchet sequence is IX-VI-V-IV, giving a clockwise rotation of
  $90^\circ$, as shown in the inset.
}
\label{fig:7}
\end{figure}

In Fig.~\ref{fig:7}(a) we plot $\langle \Delta X\rangle$ and
$\langle \Delta Y\rangle$ versus $F^{ac}_x$ for the system in
Fig.~\ref{fig:4} under $x$ direction driving
with $\alpha_{m}/\alpha_{d} = 10.0$.
A series of ratchet types appear, and there is
a double reversal in
$\langle \Delta Y\rangle$ from negative to positive
and then back to negative,
as well as in $\langle \Delta X\rangle$, which transitions from
negative to  positive and back to negative.
The resulting ratchet sequence is
IX-VI-V-IV-III-II-I-VIII-VII-VI, showing that the flow  undergoes clockwise rotation
through all the possible ratchet types or a rotation of $360^\circ$ in the inset
of Fig.~\ref{fig:7}(a).
For $y$ direction driving, $F^{ac}_y$,
Fig.~\ref{fig:7}(b) shows
that the ratchet sequence is
IX-VI-V-IV, giving a clockwise rotation of $90^\circ$.

From the ratchet behavior shown in Figs.~\ref{fig:6} and \ref{fig:7}, we can describe
the direction of ratchet motion in terms of a vector
with an amplitude of
$R=|\langle \Delta X\rangle^2 + \langle \Delta Y\rangle^2|^{1/2}$
and an orientation of $\theta$.
This ratchet vector rotates as the parameters of the system are changed, and
it can in principle point along any direction $\theta$ in the $x-y$ plane even though
the asymmetry of the substrate exists only along the $x$-direction.
This represents a new type of ratchet that arises due to the skyrmion
Hall angle,
which depends on both $\alpha_{m}/\alpha_{d}$ and drive amplitude as shown in
Fig.~\ref{fig:2}.
For the parameters we consider, increasing the ac drive or
the ratio $\alpha_m/\alpha_d$
increase the skyrmion Hall angle in the clockwise direction.

\begin{figure}
\includegraphics[width=3.5in]{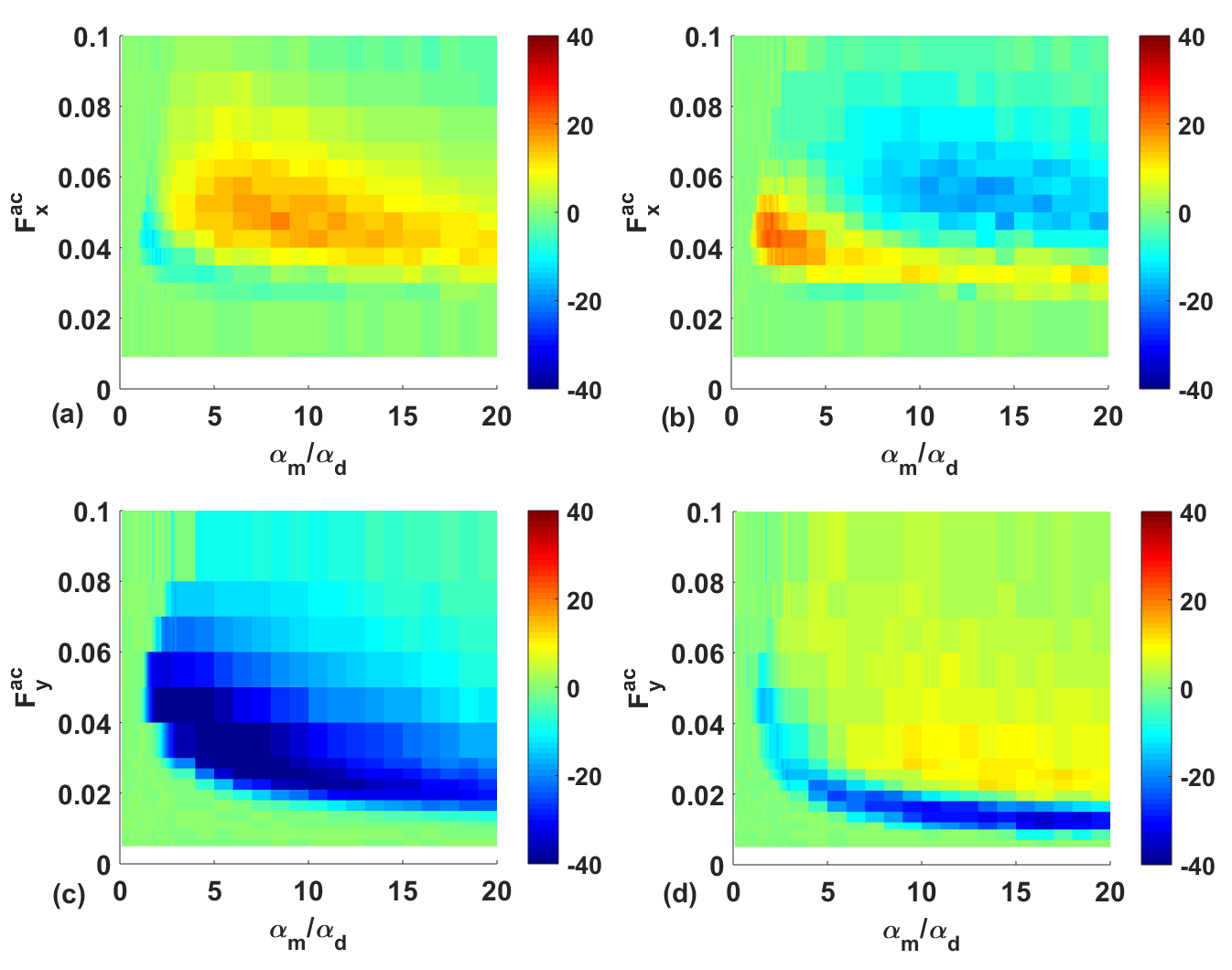}
\caption{Heat maps of (a) $\langle \Delta X\rangle$ and (b)
  $\langle \Delta Y\rangle$ as a function of
  $F^{ac}_x$ vs $\alpha_{m}/\alpha_{d}$ for the conformal array.
  Here there are ratchet reversals in both the $x$ and $y$ directions.
  (c) Heat map of $\langle \Delta X\rangle$ 
  for driving in the $y$-direction where the drift is always in
  the negative $x$ direction.
  (d) The corresponding $\langle \Delta Y\rangle$ as a function of
  $F^{ac}_y$ vs $\alpha_m/\alpha_d$
  showing a reversal.  
}
\label{fig:8}
\end{figure}

In order to get a better understanding of the evolution of the ratchet flow in
Figs.~\ref{fig:6} and \ref{fig:7},
in Fig.~\ref{fig:8} we show a heat map of the direction and magnitude of
the net flux
for $x$ direction ac driving
based on the value of $\langle \Delta X\rangle$ (Fig.8(a))
and $\langle \Delta Y\rangle$ (Fig.8(b)) after 400 ac cycles as a function of
$F^{ac}_x$ vs $\alpha_{m}/\alpha_{d}$. Here
for $F^{ac}_x < 0.3$ and $F^{ac}_x > 0.9$,
the ratchet effect is weak or absent.  It is also clear
that a reversal occurs in
both $\langle \Delta X\rangle$ and $\langle \Delta Y\rangle$
as functions of $\alpha_{m}/\alpha_{d}$ and $F^{ac}_x$.
Figure~\ref{fig:8}(d,e) shows similar heat maps for $y$ direction ac driving.
In this case the maximum intensity of the ratchet effect is stronger
and $\langle \Delta X\rangle$ is always negative, while there
is a reversal in $\langle \Delta Y\rangle$.
Since $\langle \Delta X\rangle$ is always negative, the ratchet sequence is limited
to types
III-IV-V-VI-VII.

\section{Skyrmion Density Dependence and Commensuration Effects}

\begin{figure}
\includegraphics[width=3.5in]{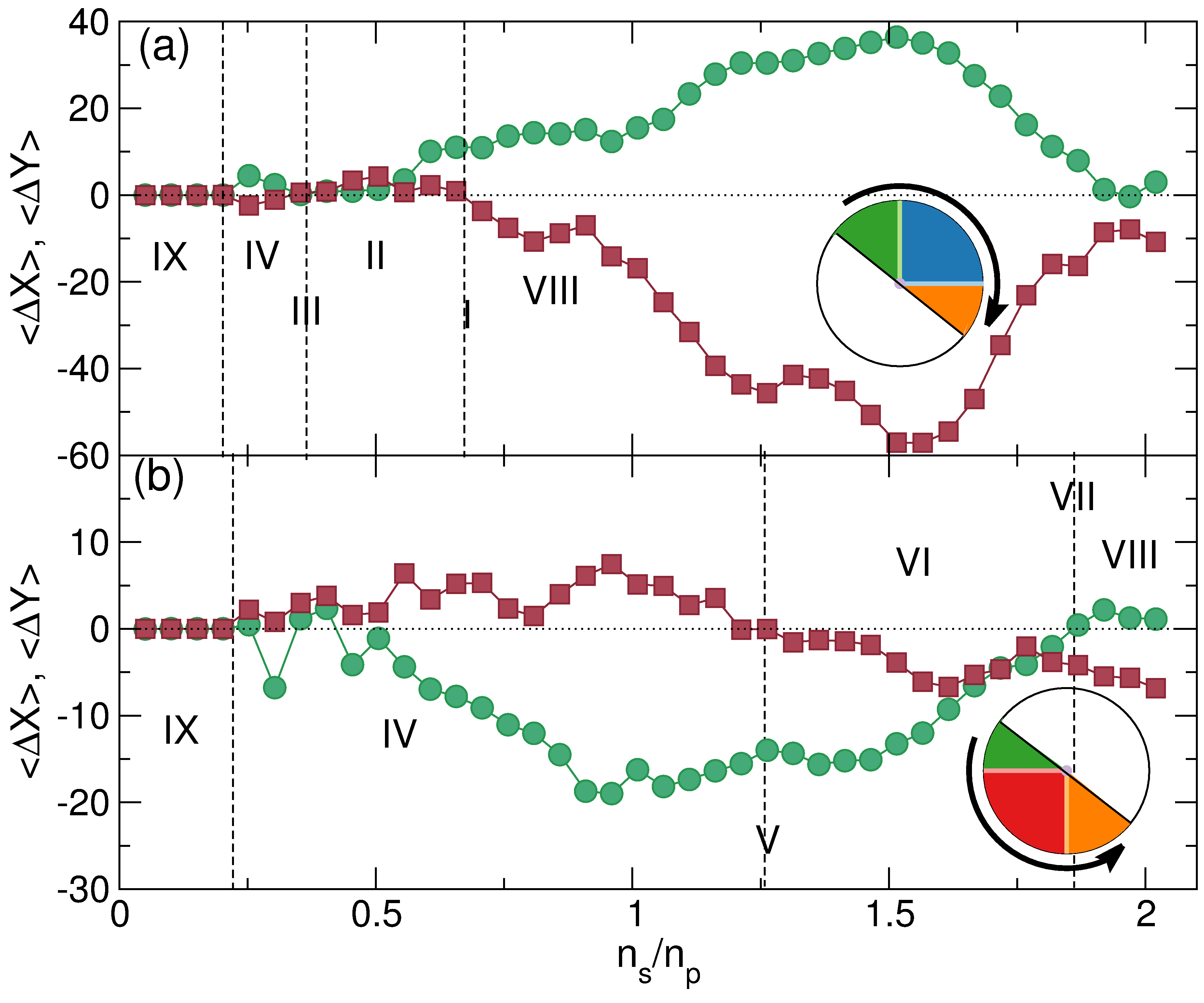}
  \caption{$\langle \Delta X\rangle$ (green circles) and
    $\langle \Delta Y\rangle$ (red squares) after 400 ac cycles vs skyrmion density $n_s/n_p$
    for $F_p=0.1$, $F^{ac} =0.05$, and
    $\alpha_{m}/\alpha_{d} = 9.962$.
    (a) Driving in the $x$ direction, $F^{ac}_x$.
    (b) Driving in the $y$ direction, $F^{ac}_y$.
    }
  \label{fig:9ab}
  \end{figure}

\begin{figure}
  \includegraphics[width=3.5in]{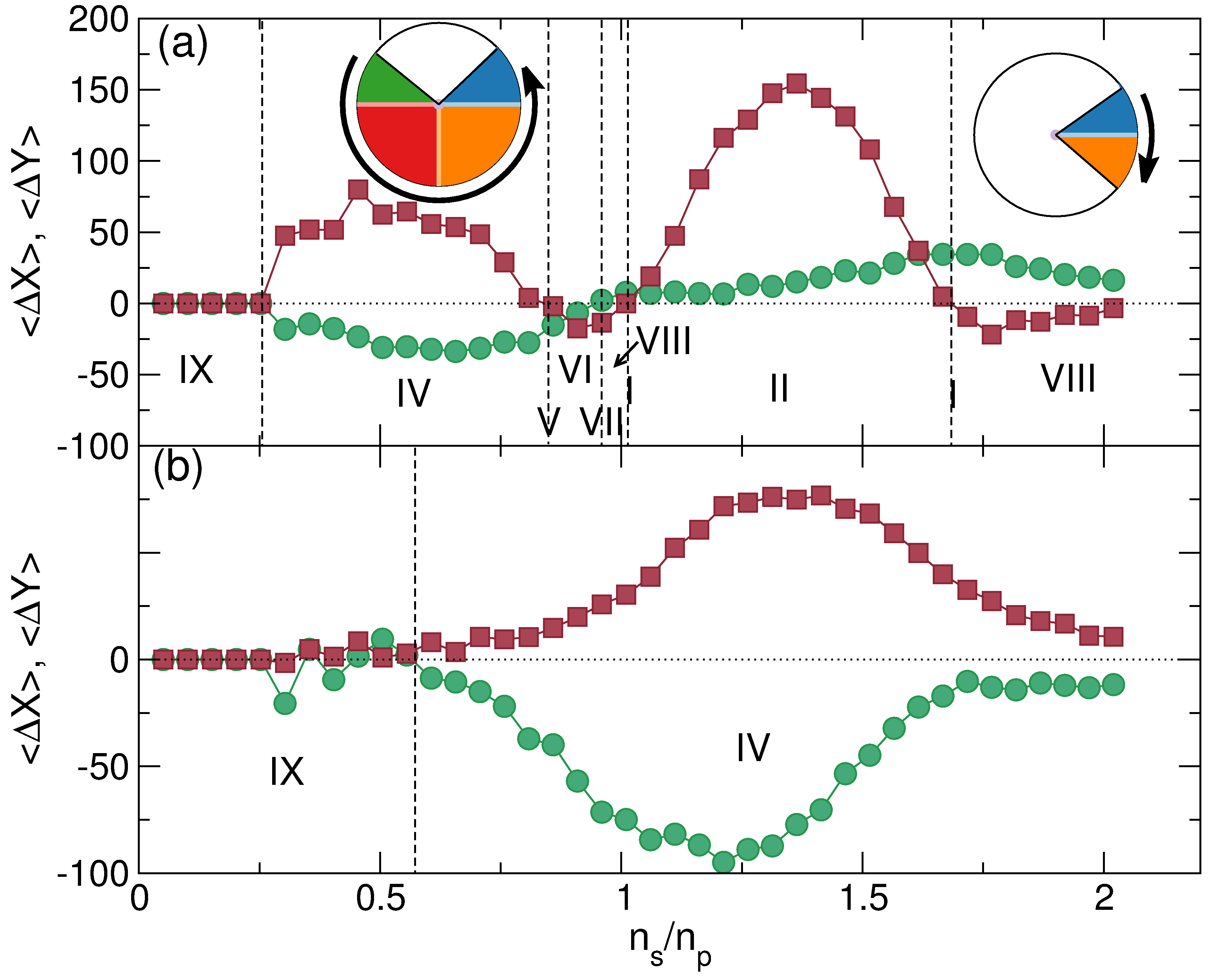}
  \caption{
$\langle \Delta X\rangle$ (green circles) and
    $\langle \Delta Y\rangle$ (red squares) after 400 ac cycles vs skyrmion density $n_s/n_p$
    for $F_p=0.5$, $F^{ac} =0.25$ and
    $\alpha_{m}/\alpha_{d} = 9.962$.
    (a) Driving in the $x$ direction, $F^{ac}_x$.  The
    ratchet flow direction initially rotates counterclockwise, followed by a clockwise
    rotation for $n_s/n_p>1.3$.
    (c) Driving in the $y$ direction, $F^{ac}_y$.
}
\label{fig:9cd}
\end{figure}

We next consider the effect of varying the skyrmion density
for a fixed pinning site density of
$n_{p} = 0.3$
at $\alpha_{m}/\alpha_{d} = 9.962$.
In Fig.~\ref{fig:9ab}(a) we plot
$\langle \Delta X\rangle$ and $\langle \Delta Y\rangle$
after 400 ac cycles for $x$ direction driving of $F^{ac}_x = 0.05$ with  $F_p=0.1$ 
over the range $0 < n_{s}/n_{p} < 2.0$.
There is a strong type VIII ratchet flux  $n_{s}/n_{p} > 0.7$,
and the ratchet sequence
IV-III-II-I-VIII
progresses clockwise around the diagram in the inset
of Fig.~\ref{fig:9ab}(a).
In general we do not observe any ratchet motion
in the single skyrmion limit of $n_s/n_p \approx 0$, indicating that
the skyrmion ratchet motion on the conformal array
is a collective effect,
unlike the ratchet effect observed
for a single skyrmion on a quasi-one-dimensional asymmetric substrate \cite{38}.
There is a weak
dip in the ratchet flux
at $n_{s}/n_{p} = 1.0$,
and the maximum ratchet flux occurs near $n_{s}/n_{p} = 1.5$,
above which the flux decreases again.
In general, the ratchet flux diminishes
for large $n_{s}/n_{p}$ where the skyrmions form
a stiff lattice that only weakly couples to the substrate.
Similar effects appear in a superconducting vortex system
for the ratchet flux at high
vortex densities in the presence of a conformal pinning array \cite{39}.
In Fig.~\ref{fig:9ab}(b) we show the same system
with
$y$ direction driving of $F^{ac}_y=0.05$.
For $n_{s}/n_{p} < 0.5$, the data is fairly noisy, but for $n_s/n_p>0.5$,
ratchet flow occurs
in both $\langle \Delta X\rangle$ and $\langle \Delta Y\rangle$ with a ratchet
sequence of
IX-IV-V-VI-VII-VIII, indicating a counter-clockwise rotation of the flow
by $180^{\circ}$  as indicated in the inset.

In Fig.~\ref{fig:9cd}(a)
we show the same system as in Fig.~\ref{fig:9ab}(a)
driven in the $x$-direction with a
pinning strength of $F_p  = 0.5$ and an ac amplitude of $F^{ac}_x=0.25$ that
have both been increased by a factor of five.
In this case, the net ratchet flux is up to 
$3.75$ times larger than
that produced when $F^{ac}_x = 0.05$ and $F_p=0.1$.
Here, $\langle \Delta Y\rangle$
is generally larger than $\langle \Delta X\rangle$,
and there are multiple reversals in the $y$ direction motion as
well as one reversal in the $x$ direction motion.
The ratchet sequence is
IX-IV-V-VI-VII-VIII-I-II for $n_s/n_p<1.25$, giving
a counter-clockwise rotation of the flow
direction by $270^{\circ}$ as shown in the leftmost inset of Fig.~\ref{fig:9ab}(a),
while for $n_s/n_p>1.25$, the ratchet sequence is
II-I-VIII, giving a clockwise rotation of $90^\circ$ as shown in the rightmost inset.
This indicates
that it is also possible to have reversals in the
direction of the ratchet flow rotation,
leading to what we term a reversible vector ratchet.
Near $n_{s}/n_{p} = 1.0$, the ratchet flux is strongly
reduced due to enhanced pinning from a commensuration effect with the
underlying substrate.
In Fig.~\ref{fig:9cd}(b), we show the ratchet flux in the same system
for driving in the $y$-direction with $F^{ac}_y=0.25$.
There is a strong type IV ratchet effect with a maximum flux
near
$n_{s}/n_{p} = 1.25$.
These results show that the skyrmion ratchet effect
is robust over a wide range of skyrmion densities, ac drive amplitudes, and
$\alpha_{m}/\alpha_{d}$ ratios.

\begin{figure}
\includegraphics[width=3.5in]{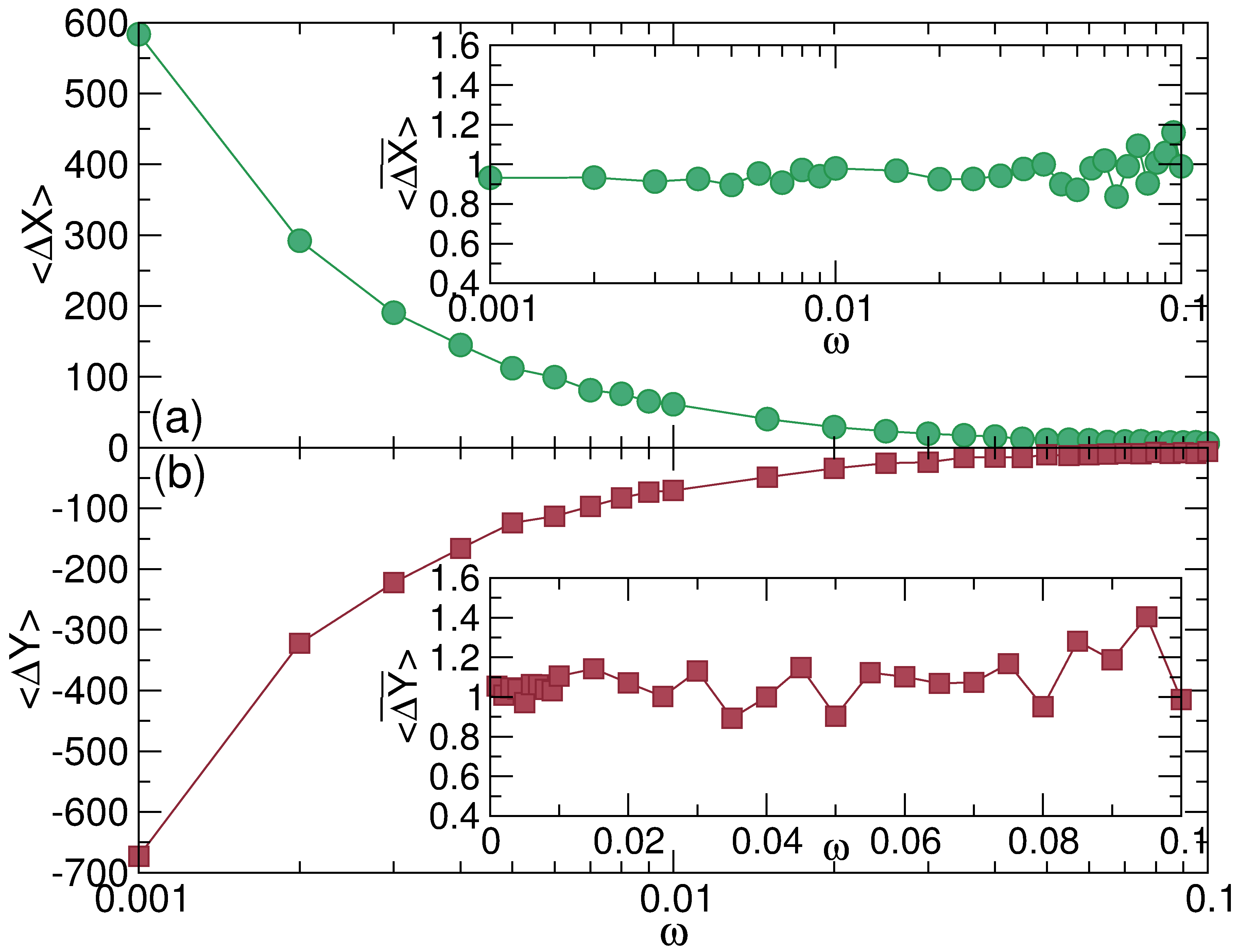}
  \caption{(a) $\langle \Delta X\rangle$
    (green circles) and
    (b) $\langle \Delta Y\rangle$ (red squares) after 400 ac cycles
    vs ac frequency $\omega$
    in samples with $\alpha_{m}/\alpha_{d} = 9.962$
    and $F^{ac}_x = 0.05$.
    In both cases, the ratchet flux
    decreases with increasing ac drive frequency.
    The insets show normalized values (a) $\langle \overline{ \Delta X}\rangle$
    and (b) $\langle \overline{\Delta Y}\rangle$ vs $\omega$.
    Normalization is achieved by dividing by the total time required
    to perform 400 ac drive
    cycles at each frequency, and then dividing by the value
    at $\omega = 0.04$.
}
\label{fig:10}
\end{figure}

We have also examined the effect of varying the ac driving frequency.
In Fig.~\ref{fig:10}(a,b) we plot
$\langle \Delta X\rangle$ and $\langle \Delta Y\rangle$ after 400 ac cycles versus
ac frequency $\omega$
in samples with $n_{s}/n_{p} = 0.3$
and $\alpha_{m}/\alpha_{d} = 9.962$.
The ratchet flux drops
with increasing $\omega$, in agreement
with observations made in overdamped systems \cite{39}.
In the insets of Fig.~\ref{fig:10}(a,b), we show the
normalized quantities
$\langle \overline{\Delta X}\rangle=\langle \Delta X\rangle/X_0\tau(\omega)$
and
$\langle \overline{\Delta Y}\rangle=\langle \Delta Y\rangle/Y_0\tau(\omega)$,
where
$\tau(\omega)$ is the number of simulation time steps required to complete 400 ac drive
cycles at a driving frequency $\omega$, 
$X_o$ is the value of $\langle \Delta X\rangle/\tau(\omega)$ at $\omega=0.04$, and
$Y_o$ is the value of $\langle \Delta Y\rangle/\tau(\omega)$ at $\omega=0.04$.
The normalized measures indicate that the net ratchet flux
remains roughly constant when adjusted for the amount of time spent ratcheting
at the different ac drive frequencies.

\section{Particle Trajectories}

\begin{figure}
\includegraphics[width=3.5in]{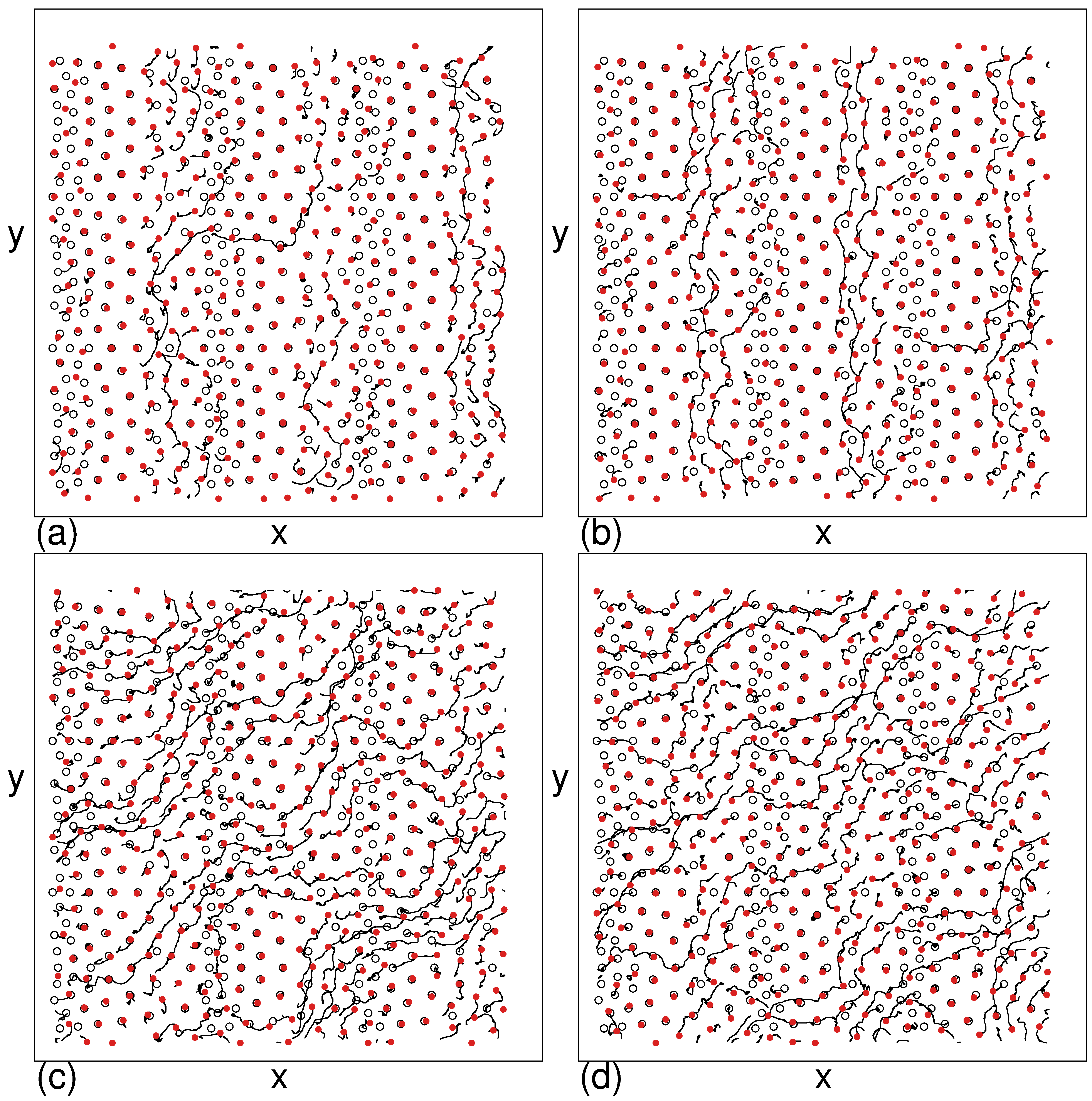}
\caption{ Skyrmion positions (filled dots),
  pinning site locations (open circles), and trajectories (lines) for
  $y$ direction ac driving $F^{ac}_y$
  in a sample with $n_{s}/n_{p} = 1.0$.
  (a) The positive ac drive cycle for $F^{ac}_y = 0.013$.
  (b) The negative ac drive cycle for $F^{ac}_y=0.013$.  At this drive, a type VI
  ratchet with motion in the negative $y$ and negative $x$ directions occurs.
  (a) The positive ac drive cycle for $F^{ac}_y=0.03$.
  (a) The negative ac drive cycle for $F^{ac}_y=0.03$.
  Here, there is a type IV ratchet with motion in the negative $x$ and positive $y$ directions.
}
\label{fig:11}
\end{figure}

We image the skyrmion trajectories on either side of a ratchet reversal
in order to understand how the geometry of the pinning array affects the
skyrmion motion and how the amplitude of the ac drive can change the direction of the
net ratchet flux.
In Fig.~\ref{fig:11}(a,b) we plot the skyrmion positions, pinning site locations, and
skyrmion trajectories  in a sample with
$\alpha_{m}/\alpha_{d} = 9.962$
under a $y$ direction ac drive of
$F^{ac}_y=0.013$, which produces a type VI ratchet with strong flux in the negative $y$
direction and weak flux in the negative $x$ direction.
During the positive portion of the ac drive cycle, shown in Fig.~\ref{fig:11}(a),
the skyrmions predominantly
move  in the positive $y$ direction.
The flow is concentrated in
the regions of lower pinning density, and there is a small amount of skyrmion hopping
in the positive $x$ direction, which is the hard flow direction of the substrate
asymmetry.
If the pinning sites were not present, during the positive portion of the ac drive cycle the
skyrmions would move with a Hall angle of
$85^{\circ}$ relative to the positive $y$ axis.
Instead, in Fig.~\ref{fig:11}(a), the Hall angle is nearly zero since skyrmion motion
in the positive $x$ direction is blocked by the regions of dense pinning.
The Magnus term couples the $x$ and $y$ motion and causes the positive $y$ direction
to act like a hard flow direction
even though there is no asymmetry in the substrate
along the $y$ direction.
During the negative portion of the ac drive cycle, illustrated in Fig.~\ref{fig:11}(b),
the motion is mostly in the negative $y$ direction, with some hopping in the negative
$x$ direction.  Since the negative $x$ direction is the easy flow direction of the ratchet
asymmetry, the Magnus coupling causes the negative $y$ direction to act like an easy
flow direction, and the net ratchet flux during the entire cycle is larger in the negative $y$
direction than in the positive $y$ direction, producing a net negative $y$
and negative $x$ flow.
Figure~\ref{fig:11}(c) shows the positive portion of the ac cycle for a drive of
$F^{ac}_y=0.03$, while Fig.~\ref{fig:11}(d) shows the negative portion of the ac cycle
at the same drive.  For this ac drive amplitude,
there is a strong ratchet flux
in the negative $x$ direction and a weaker
ratchet flux in the positive $y$ direction, giving a type IV ratchet effect.
The ac drive is strong enough that, during the positive
portion of the ac cycle in Fig.~\ref{fig:11}(c),
the skyrmions can pass through the densely pinned regions,
and the resulting Hall angle is larger than that observed at the lower ac amplitude of
$F^{ac}_y=0.013$.
During the negative
portion of the ac cycle, shown in Fig.~\ref{fig:11}(d),
the skyrmions continue to pass through the densely pinned regions, but since
the negative $x$ direction is the easy flow direction of the substrate asymmetry, the
net amount of negative $x$ motion is increased compared to
that which occurs during the positive portion of
the ac cycle, and correspondingly the amount of motion in the negative $y$ direction
is decreased.
Thus, for fixed pinning strength and skyrmion
density,
the ratchet flow rotates with increasing ac amplitude $F^{ac}_y$ due to the depinning
process in the $x$ direction and the
increasing
Hall angle, as shown in Fig.~\ref{fig:1}.

\begin{figure}
\includegraphics[width=3.5in]{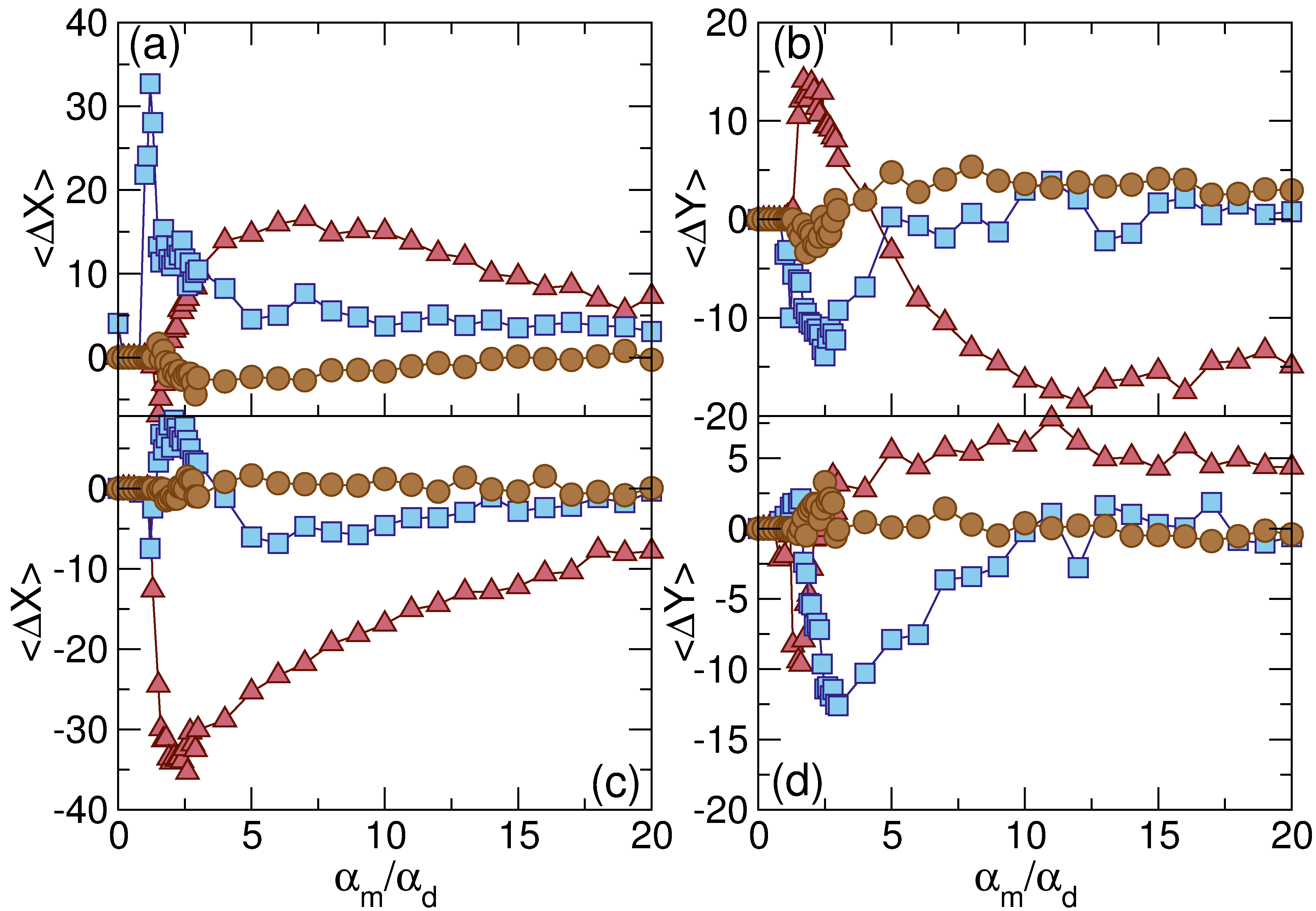}
\caption{
  Ratchet motion in the different arrays illustrated in Fig.~\ref{fig:1}:
  conformal pinning array (red triangles),
  square gradient array (blue squares), and random gradient array (brown circles),
  in samples with $n_s/n_p=0.3$.
  (a) $\langle \Delta X\rangle$ and (b) $\langle \Delta Y\rangle$ 
  after 400 ac cycles vs $\alpha_{m}/\alpha_{d}$ for $x$ direction driving $F^{ac}_x$.
  (c) $\langle \Delta X\rangle$ and (d) $\langle \Delta Y\rangle$
  after 400 ac cycles vs $\alpha_{m}/\alpha_{d}$ for $y$ direction driving $F^{ac}_y$.
}
\label{fig:12}
\end{figure}

\section{Square and Random Gradient Arrays}
In Fig.~\ref{fig:12}(a,b) we show
$\langle \Delta X\rangle$ and $\langle \Delta Y\rangle$ after 400 ac cycles
versus $\alpha_{m}/\alpha_{d}$
in samples with
$n_{s}/n_{p} = 0.3$
containing either the square gradient array illustrated in Fig.~\ref{fig:1}(b) or the
random gradient array shown in Fig.~\ref{fig:1}(c).  Also shown for comparison is a
sample with a conformal array.
Here the square gradient array produces a large ratchet flux
for low $\alpha_{m}/\alpha_{d} < 5.0$, and 
in some cases the flow is in the opposite direction to
that observed in the conformal array.
The random gradient array in general shows a much smaller ratchet
flux that is primarily in the negative $x$ and positive $y$ directions,
which is opposite to the flux observed for the conformal array.
Figure~\ref{fig:12}(c) shows $\langle \Delta X\rangle$
vs $\alpha_{m}/\alpha_{d}$ for the same systems under $y$ direction
ac driving, $F^{ac}_y$.
In this case, the conformal array always produces a negative $x$ ratchet flux,
while the square gradient array shows a weaker ratchet flux as well as a reversal
from positive $x$ to negative $x$ flow
near $\alpha_{m}/\alpha_{d} = 5.0$.
The random gradient array does not show any appreciable ratchet flux.
In Fig.~\ref{fig:12}(d), the corresponding
$\langle \Delta Y\rangle$ versus $\alpha_{m}/\alpha_{d}$ plot indicates that
the ratchet flux of the square gradient array is comparable to
or even higher than that of the conformal array for $\alpha_m/\alpha_d<5$, while the
random gradient array shows almost no ratchet flux.
We observe similar effects for 
fixed $\alpha_{m}/\alpha_{d}$ and varied ac amplitude  $F^{ac}$.
In general, the conformal array produces the largest ratchet flux, while the ratchet
flux for the square gradient array is weaker, and that of the random gradient array
is the weakest.

\section{Summary}
We have shown that ac driven skyrmions interacting with
two-dimensional gradient pinning arrays
represent a realization of a new type of ratchet system that we
call a vector ratchet.
In overdamped systems, the ratchet flux is limited to flowing parallel to the substrate
asymmetry direction in the forward or reverse direction.
In contrast, the strongly non-dissipative Magnus term found in skyrmion systems produces
a skyrmion Hall angle that couples the motion parallel and perpendicular to the substrate
asymmetry direction.
The resulting dc ratchet drift
generated by the ac drive can be described as a vector
which can rotate counter-clockwise or clockwise in the
$x-y$ plane as the ac amplitude or the ratio of the
Magnus term to the dissipative term is varied,
so that it is possible to realize reversals in the ratchet
flux in both the $x$ and $y$ directions.
We show that this vector ratchet appears for ac driving both parallel to and
perpendicular to the substrate asymmetry direction.
The ratchet
reversals we observe are a result of collective skyrmion interactions, as previous work
on individual skyrmions interacting with asymmetric substrates showed no ratchet
reversals.
In addition to reversals in the ratchet flux in the $x$ and $y$ directions,
the angular rotation of the ratchet vector itself
can also show a reversal.
We find that it is possible to have rotations
of the ratchet vector of up to $360^{\circ}$, indicating that
vector ratchets can be used to direct skyrmion motion in any
in-plane direction.
Thus, the vector ratchet could serve as a powerful new method
for controlling skyrmion motion.
Vector ratchets should be general to systems of collectively interacting particles driven
over asymmetric substrates where Magnus type effects are present.

\acknowledgments
This work was carried out under the auspices of the 
NNSA of the 
U.S. DoE
at 
LANL
under Contract No.
DE-AC52-06NA25396.

\end{document}